\documentclass[aps,prb,reprint]{revtex4-2}
\usepackage{amsmath}
\usepackage{amssymb}
\usepackage{graphicx}
\usepackage{hyperref}
\begin{document}
\title{Peierls transition driven ferroelasticity in two-dimensional $d$-$f$ hybrid magnet}
\author{Haipeng You}
\author{Yang Zhang}
\author{Jun Chen}
\author{Ning Ding}
\author{Ming An}
\author{Lin Miao}
\author{Shuai Dong}
\email{Corresponding author. Email: sdong@seu.edu.cn}
\affiliation{School of Physics, Southeast University, Nanjing 211189, China}
\date{\today}

\begin{abstract}
For broad nanoscale applications, it is crucial to implement more functional properties, especially those ferroic orders, into two-dimensional materials. Here GdI$_3$ is theoretically identified as a honeycomb antiferromagnet with large $4f$ magnetic moment. The intercalation of metal atoms can dope electrons into Gd's $5d$-orbitals, which alters its magnetic state and lead to Peierls transition. Due to the strong electron-phonon coupling, the Peierls transition induces prominent ferroelasticity, making it a multiferroic system. The strain from undirectional stretching can be self-relaxed via resizing of triple ferroelastic domains, which can protect the magnet aganist mechnical breaking  in flexible applications.
\end{abstract}
\maketitle

\section{Introduction}
Since the discovery of graphene, two-dimensional (2D) materials have formed a blooming field, containing many semiconductors, semimetals, as well as topological materials, which have brought great opportunities for applications in microelectronics, optoelectronics, catalysis, etc., for their excellent physical and chemical properties. In fact, more functionalities can be implemented into 2D materials, which can provide superior performances than their three-dimensional (3D) counterparts. Very recently, 2D materials with intrinsic ferroic oders (e.g. magnetism or polarity) became an emerging branch of the 2D family, which are of great interests for the spintronic/electronic applications in nanoscale \cite{Gong:Sci,An:APLM,Wu:Wcms,Tang:JPCL,Zhou:Fop}.

Currently, most studied ferromagnetic (FM) and antiferromagnetic (AFM) 2D materials originate from van der Waals (vdW) or MXene layered materials with transition metals  (e.g. V, Cr, Mn, Fe, Co, Ni, Cu) \cite{McGuire:Crystal,Zhang:Jacs,Hidalgo:PRB,McGuire:PRB,Sun:PRB,Ding:PRB,Kim:PRL,Huang:PRL,Yu:Fop,Liao:CPL,Coak:npjqm,you:pccp}, i.e., with $3d$-electron spins. Besides, $4f$ electrons, with even stronger Hubbard correlation and more spatially localization, can also provide magnetic moments, but have been much less studied in 2D materials. The $4f$-electron magnets have some unique physical characteristics. For example, the spin-orbit coupling (SOC) is typically much stronger for $4f$-orbitals, which may lead to larger magnetocrystalline anisotropy, an important condition to stablize magnetic orders in the 2D limit. And the maximum local spin moment can be larger than the $3d$ one. However, due to the localized distribution of $4f$ orbitals, the exchanges between neighbor $4f$ spins are typically weaker, leading to lower ordering temperatures.

Recently, a 2D monolayer with rare earth metal, GdI$_2$, was predicted to be a FM semiconductor with Curie temperature ($T_{\rm C}$) close to room temperature \cite{Wang:MH}. Its exotic large exchange interaction comes from the $4f^7$+$5d^1$ hybridization of Gd$^{2+}$. The spatially expanded $5d$ electron play as a bridge to couple localized $4f$ spins. This work opened a door to pursuit high-performance 2D magnets based on $4f$ spins.

Here another 2D $f$-electron halide, GdI$_{3}$ monolayer, is theoretically studied. Different from the compact triangular structure of GdI$_2$, GdI$_{3}$ monolayer owns the honeycomb geometry of Gd$^{3+}$ spins. The ``vacancies" within the hexatomic rings can host dopants, providing an efficient route to tune its physical properties. Indeed, here the magnetic ground state can be tuned by electron doping, accompanying the Peierls transition. Prominent ferroelasticity is induced by this Peierls transition. Thus, the doped GdI$_{3}$ becomes a 2D multiferroic system with superior elasticity, adding additional value to the 2D magnetic materials.

\section{Methods}
The first-principles calculations are performed based on the spin-polarized density functional theory (DFT) implemented in Vienna {\it ab initio} Simulation Package (VASP) code \cite{Kresse:Prb}. The projected augmented wave (PAW) method is used to describe the ion-electron interaction, and the kinetic energy cutoff for the plane-wave is set as $500$ eV. The Perdew-Burke-Ernzerhof (PBE) parametrization of the generalized gradient approximation (GGA) is used for the exchange-correlation functional \cite{Perdew:PRL}. The Hubbard correlation is considered using the rotationally invariant GGA+$U$ method introduced by Liechtenstein \textit{et al.} \cite{Liechtenstein:Prb}, with $U$=$9.2$ eV and $J$=$1.2$ eV applied on Gd's $4f$ orbitals \cite{Larson:Prb}.

The structures are relaxed with the conjugate gradient method until the Hellmann-Feynman force on each atom is less than $0.01$ eV/\AA. The Monkhorst-Pack scheme is chosen to sample the Brillouin zone, with $4\times4\times1$ $k$-point for the primitive cell and $2\times2\times1$ for the supercell. For monolayers, a $20$ \AA{} vacuum layer is added to avoid interactions between adjacent layers. The vdW interaction is described by the DFT-D2 functional \cite{Grimme:JCC}. The phonon band structure is calculated using the density functional perturbation theory (DFPT) \cite{Gonze:Prb,Togo:SM}.

Spin-orbit coupling (SOC) is considered when calculating the magnetocrystalline anisotropy energy (MAE). The MAE coefficient $K$ is also estimated as $E_c-E_a$, where $E_x$ denotes the energy of ferromagnetic state with spin pointing along $x$-direction.

\section{Results \& Discussion}
\subsection{Physical properties of GdI$_3$}
\begin{figure}
\includegraphics[width=0.48\textwidth]{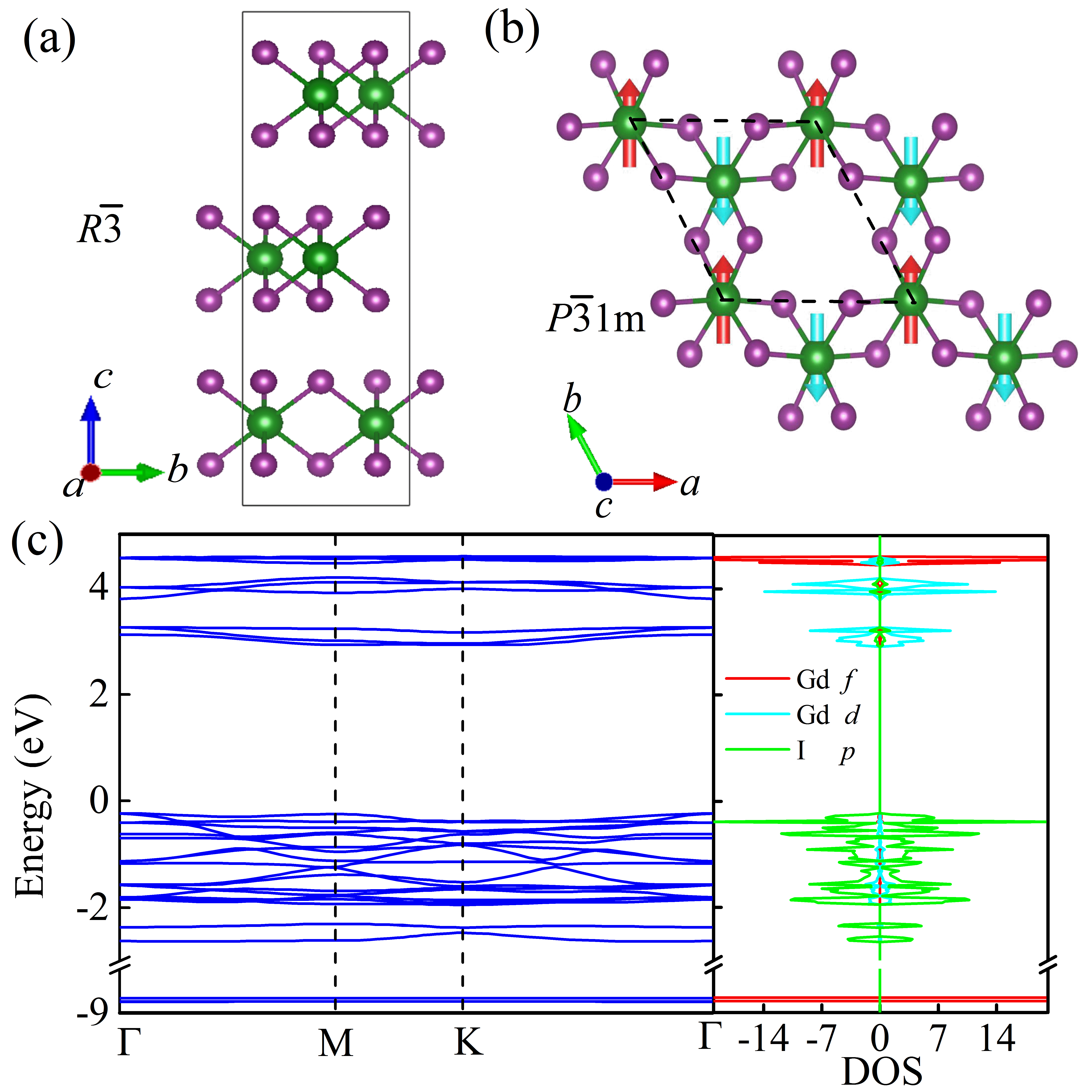}
\caption{Physical properties of GdI$_{3}$. (a) The structure of vdW layered GdI$_{3}$. (b) The top view of monolayer exfoliated from the vdW layered phase. The primitive cell is indicated by the broken lines. Green: Gd; Purple: I. The spins of N\'eel type antiferromagnetic (AFM) ground state are indicated by arrows. (c) The band structure and density of states (DOS) of GdI$_{3}$ monolayer. The lowest conducting band is contributed by Gd's $5d$ orbitals, while the highest valence band is from I's $5p$ oribitals.}
\label{fig1}
\end{figure}

For GdI$_{3}$ bulk, there are two possible structures. One predicted phase for various $R$I$_3$ ($R$: rare earth like Tb, Dy, Er, etc) is consisted by vdW-coupled chains (with space group $P6_3/mmc$) \cite{TbI3}, as shown in Fig. S1 of SM \cite{SM}.  Another phase is vdW layered structure with space group $R\bar{3}$, as shown in Fig.~\ref{fig1}(a), which was synthesized experimentally more then half a century \cite{Asprey:IC}.

To clarify the ground state, both structures of $P6_3/mmc$ and $R\bar{3}$ phases are calculated for comparison. As summarized in Table~S1 \cite{SM}, our optimized lattice constants agree with the previous experiment \cite{Asprey:IC}, implying reliable results. According to our calculation, the energy of $P6_3/mmc$ phase is $366$ meV/f.u. higher than that of $R\bar{3}$ phase, implying the more stable $R\bar{3}$ phase, in consistent with experimental observation. The calculated Gd's magnetic moment are very close to the expectation value $7$ $\mu_{\rm B}$/Gd, coming from its half-filled $4f$ orbitals, while the residual moment on I ion is negligible small.

For the $R\bar{3}$ GdI$_3$ bulk with 1T-BiI$_3$-type vdW layers, the I-Gd-I sandwich layers are stacked along the $c$-axis in an ABC sequence. Each Gd ion is caged within the I octahedron, while the neighbor octahedra connect via the edge-sharing manner [Fig.~\ref{fig1}(b)], similar to the crystal structure of 1T-phase CrI$_3$ \cite{Gong:Sci}.

The GdI$_{3}$ monolayer can be exfoliated from its layered bulk. The exfoliation process is simulated  to obtain the cleavage energy, as shown in Fig.~S2 in SM \cite{SM}. Compared with the cleavage energy of graphite ($0.37$ J/m$^2$), GdI$_{3}$ owns a smaller value ($0.12$ J/m$^2$), implying its easier exfoliation. Then in the following the GdI$_3$ monolayer will be calculated.

To investigate the magnetic ground state, four most possible magnetic orders, i.e., ferromagnetic (FM) and three AFM ones (N\'eel, Zigzag, and Stripy, as shown in Fig.~S3 in SM \cite{SM}), are considered \cite{An:Jpcc}. According to our DFT calculation, the N\'eel-type AFM state [Fig.~\ref{fig1}(b)] has the lowest energy, as compared in Table~S2 in SM \cite{SM}. This  N\'eel-type AFM state is under expectation since the half-filling $4f$ orbitals prefer the AFM coupling according to the Goodenough-Kanamori rule. The energy difference between the FM and N\'eel-type AFM states is only $1.8$ meV/f.u.. Such weak exchange is reasonable considering the spatially localized $4f$ orbitals. Therefore, the expected N\'eel temperature ($T_{\rm N}$) for GdI$_3$ monolayer should be very low. The value of MAE is $-0.03$ meV/Gd, indicating an easy axis pointing out-of-plane. Such a weak magnetic anisotropy is due to the half-filled $4f$ orbitals, which cancels the effect of SOC.

The electronic structure of N\'eel-type AFM GdI$_3$ monolayer is shown in Fig.~\ref{fig1}(c). It is a Mott insulator, with fully split narrow $4f$ bands, due to large Hubbard replusion. The lowest conducting band is contributed by Gd's $5d$ orbitals, instead of $4f$'s upper Hubbard band. And the highest valence band is from I's $5p$ orbitals.

The low-temperature antiferromagnetism in GdI$_3$ is in sharp contrast with the predicted high-temperature ferromagnetism in GdI$_2$. Then it is interesting to explore the electron doping effect in GdI$_3$. Is it possible to achieve high-temperature ferromagnetism by electron doping? Is it possible to obtain metal-insulator transition with electron doping? Is there any other emergent physics in doped systems, like what happens in high-$T_{\rm C}$ superconducting cuprates and colossal magnetoresistive manganties \cite{Dagotto:Sci}?

\subsection{Peierls transition in Li intercalated GdI$_3$}
\begin{figure}
\includegraphics[width=0.48\textwidth]{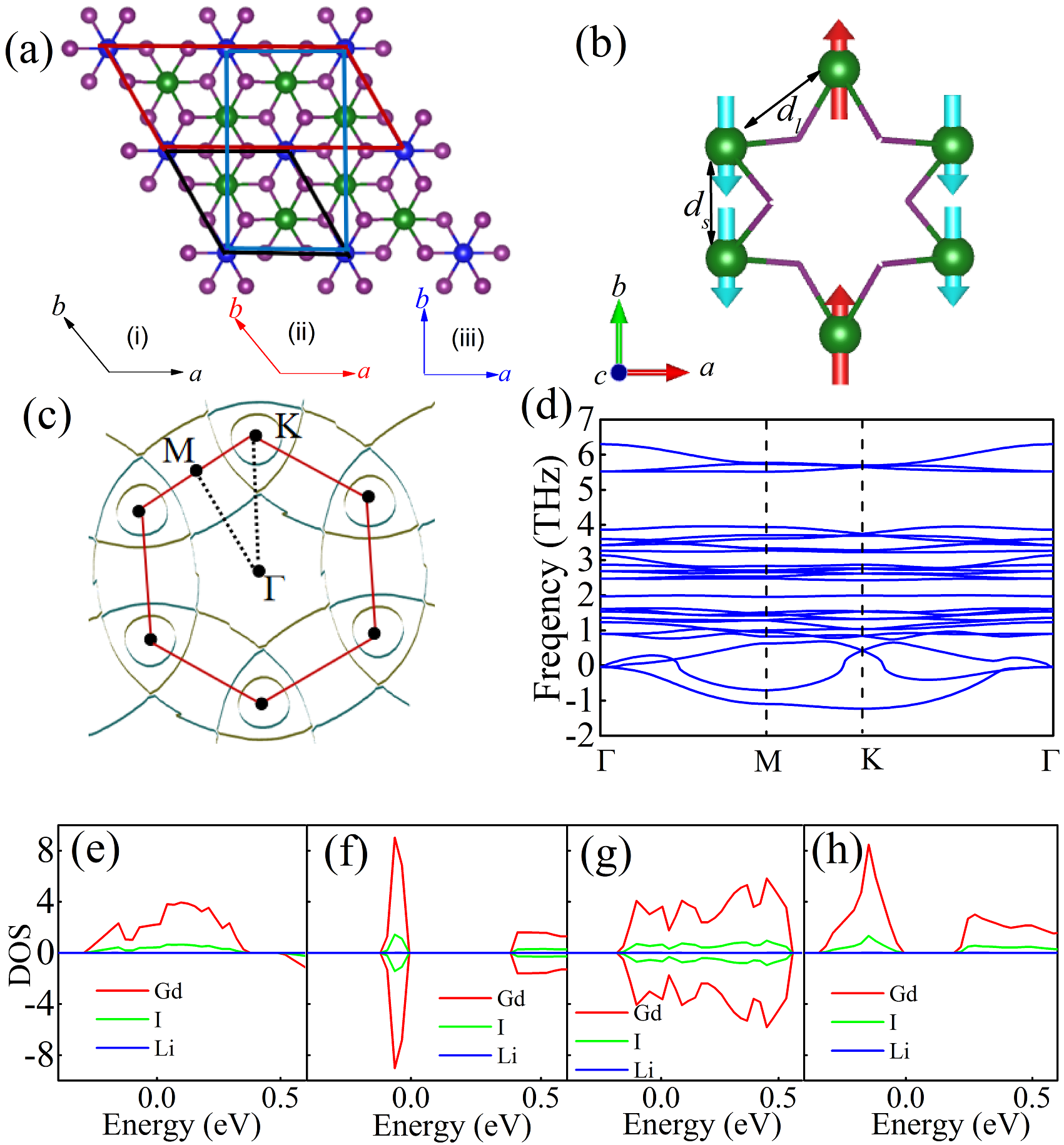}
\caption{The Peierls transition of (GdI$_3$)$_2$Li. (a) Structure of Li-implanted GdI$_3$ monolayer. Blue: Li. Black rhombus: primitive cell of FM or N\'eel AFM states; Red parallelogram: primitive cell of Zigzag AFM state; Blue rectangle: primitive cell of Stripy AFM state. (b) The Gd-framework for Stripy-AFM phase. The structural dimmerization can be visualized clearly: the shorter $d_s$ between parallel-spin Gd-Gd pairs and longer $d_l$ between antiparallel-spin ones. (c) Fermi surface of the FM $P\bar{3}1m$ state, coming from Gd's $5d$ electrons. The hexagonal Fermi surface surrounding the $\Gamma$ point provides the possibility for Fermi surface nesting. (d) Phonon spectrum of the FM $P\bar{3}1m$ state, which indicates the dynamic unstability. (e-h) Comparison of DOS's near the Fermi level. (e) FM state on optimized $P\bar31m$ structure. (f)  Stripy state on optimized $C2/m$ structure. (g) Stripy state on $P\bar31m$ structure. (h) FM state on optimized $C2/m$ structure.}
\label{fig2}
\end{figure}

To answer these questions, Li atoms are intercalated into the interstitial positions of hexatomic rings, as shown in Fig.~\ref{fig2}(a). Now the chemical formula of monolayer becomes (GdI$_3$)$_2$Li, and the valence of Gd becomes fractional $+2.5$, which would be a source for emergent phenomena.

\begin{table}
\caption{Optimized structures of (GdI$_3$)$_2$Li with different magnetic orders. Lattice constants ($a$ and $b$) and neares-neighbor Gd-Gd distances ($d_l$: longer one; $d_s$: shorter one) are in unit of \AA. The shapes of primitive cells can be found in Fig.~\ref{fig2}(a). The energies are in unit of meV/f.u. and the FM one with high symmetric structure is taken as the reference.}
\begin{tabular*}{0.48\textwidth}{@{\extracolsep{\fill}}lcccccc}
\hline \hline
Order & Space group & Energy & $a$ & $b$ & $d_l$ & $d_s$ \\
\hline
FM & $P\bar{3}1m$ & 0 & 7.502 & - & 4.331 & - \\
N\'eel & $P\bar{3}1m$ & 93  & 7.503 & - & 4.332 & - \\
Zigzag & $C2/m$ & 5.5 & 7.438 & 7.678 & 4.592 &  4.207\\
Stripe & $C2/m$ & -52.6 &  7.662 & 12.730 & 4.625 & 3.773  \\
FM & $C2/m$ &  -49.9 &  7.680  & 12.726 & 4.639 & 3.761 \\
\hline \hline
\end{tabular*}
\label{table2}
\end{table}

Then the structure of (GdI$_3$)$_2$Li is relaxed with aforementioned four magnetic orders. As compared in Table~\ref{table2}, the Stripy-AFM state [Fig.~\ref{fig2}(b)] owns the lowest energy. The magnetic moment becomes $\sim7.5$ $\mu_{\rm B}$/Gd, as expected. The MAE coefficient $K$ for (GdI$_3$)$_2$Li is estimated as $0.46$ meV/Gd, which prefers the in-plane alignment of spins. The enhanced magnetic anisotropy is due the strong SOC of partial filled Gd's $5d$ orbitals. Meanwhile, there is a strong lattice deformation, i.e., the Peierls-type dimerization, which reduces the symmetry from trigonal to monoclinic $C2/m$ (No. 12). There is strong disproportion of nearest-neighbor Gd-Gd distance $d$, which is seriously split into two types: longer $d_l$ and shorter $d_s$, as indicated in Fig.~\ref{fig2}(b) and listed in Table~\ref{table2}.

This Stripy antiferromagnetism and strong lattice deformation is nontrivial. A direct question is whether the lattice deformation is driven by the magnetostriction effect since the Stripy antiferromagnetism is also dimerized, or in reverse the Stripy antiferromagnetism is driven by the lattice dimerization? To clarify the real origin,  the Fermi surface and phonon spectrum of high-symmetric FM state is analysized. As shown in Fig.~\ref{fig2}(c), there are small pockets around the $K$ point. In addition, six large ``circles'' cross with each others and form  a hexagon surrounding the $\Gamma$ point, which provides the possibility for Fermi surface nesting, i.e., the electronic driving force for Peierls transition.

Accordingly, for the FM state with high symmetric structure, the imaginary frequencies of phonon spectrum [Fig.~\ref{fig2}(d)] also suggest the spontaneous structural unstability. In fact, the optimized $C2/m$ structure with FM order leads to even slightly larger disproportion between  $d_l$ and $d_s$, as listed in Table~\ref{table2}. Thus, the appearence of lattice dimerization is not driven by the magnetostriction of Stripy antiferromagnetism, but due to the instability of electronic structure.

\begin{figure}
\includegraphics[width=0.48\textwidth]{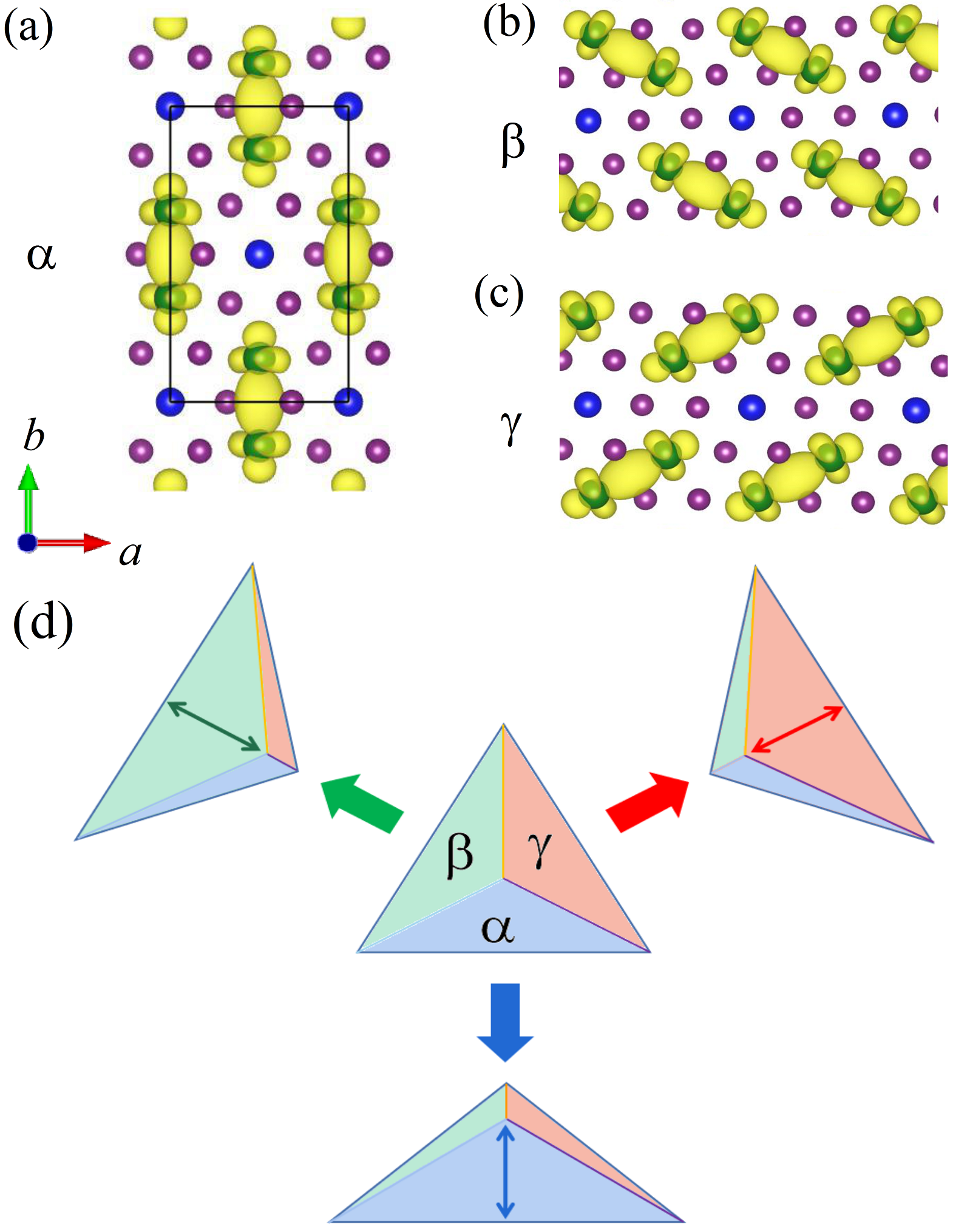}
\caption{(a) Schematic of electron distribution of Gd's $5d$ electrons. The electron spindles are along the $b$-axis in this $\alpha$ domain. The associated ferroelasticity can be characterized by the $b/a$ ratio of black rectangle, which is $1.661$, $\sim4\%$ shorter than the original $1.732$ for the $P\bar31m$ phase. (b-c) Two other ferroelastic domains with electron spindles along different directions, named as $\beta$ and $\gamma$ domains. (d) Origin of super-elasticity. The macroscopic shape of (GdI$_3$)$_2$Li monolayer can be deformed by external forces. The inner strain can be relaxed to some extent by the rotation of electron spindles, i.e. the resize of corresponding ferroelastic domains.}
\label{fig3}
\end{figure}

The Peierls-type dimerization can tune the band structure, as compared in Figs.~\ref{fig2}(e-h). The FM state with optimized $P\bar31m$ structure is metallic [Fig.~\ref{fig2}(e)], while the Stripy state with optimized $C2/m$ structure is a semiconductor. To futher clarify the origin of gap openning, the Stripy-AFM order is imposed on the $P\bar31m$ structure, which can not open the band gap [Fig.~\ref{fig2}(g)]. In contrast, the FM order on the $C2/m$ structure  can open a band gap at the Fermi level [Fig.~\ref{fig3}(h)].

Thus, it is unambiguous to conclude that the physical properties of (GdI$_3$)$_2$Li, including the lattice deformation, magnetic transition, and band gap openning, are all dominated by the  Peierls-type transition. On one hand, with both optimized $C2/m$ structures, the energy difference between FM and Stripy AFM states are so small ($2.7$ meV/Gd higher for the FM state), implying low N\'eel temperature. Considering the large magnetic moment ($7.5$ $\mu_{\rm B}$/Gd), the Zeeman energy gained from a moderate magnetic field $\sim6.2$ T can overcome the small energy difference and drive an AFM-FM phase transition. At finite temperature, the required magnetic field can be even lower with the help of thermal fluctuation. This magnetic transition can lead to negative magnetoresistive, considering the different band gaps of FM phase ($0.19$ eV) and Stripy-AFM phase ($0.38$ eV), as compared between Fig.~\ref{fig2}(f) and \ref{fig2}(h). On the other hand, the energy gain from lattice dimerization reaches $49.9$ meV/Gd with the FM order, implying that the lattice dimerization should occur at a much higher temperature than the N\'eel temperature of Stripy-AFM order.

Beside the magnetic transition, the Peierls transition can lead to an even more interesting physical phenomenon, i.e., the ferroelasticity \cite{Wu:Nl}. As stated before, the driving force of Peierls transition is Gd's $5d$ electron, which can form an undirectional charge density wave, as shown in Fig.~\ref{fig3}(a). The $5d$ electrons mostly stay in the middle of shorter Gd-Gd pair, while the middle of longer Gd-Gd pair is almost empty, i.e., the bond-centered charge ordering \cite{Efremov:Nm}. These electron spindles break the three-fold rotational symmetry, induces a shrunk lattice constant along the spindle direction, i.e. the $b$-axis of  $\alpha$ domain as shown in Fig.~\ref{fig3}(a). This ferroelastic deformation can reach $4\%$, comparable with the tetragonality of tetragonal BaTiO$_3$. The coexisting of ferroelasticity and antiferromagnetism makes (GdI$_3$)$_2$Li a multiferroic system, although it does not contain ferroelectricity.

Furthermore, the three-fold rotational symmetry of parent phase allows other two ferroelastic domains with different orientations of electron spindles, i.e., the $\beta$ \& $\gamma$ domains, as shown in Fig.~\ref{fig3}(b-c).  Starting from an anisotropic state with same amounts of $\alpha$-$\beta$-$\gamma$ domains [center of Fig.~\ref{fig3}(d)], external forces along certain directions can tune the detailed balance among these triple ferroelastic domains. Then the change of ferroelastic domains can relax the inner strain by rotating these electron spindles, making the (GdI$_3$)$_2$Li monolayer more elastic against mechanical damage in certain range, which is crucial for flexible applications. Very recently, an experiment demonstrated the super-elasticity of free-standing ferroelectric BaTiO$_3$ membrane, which was attributed to the continuous electric dipole rotation \cite{Dong:Sci}. In fact, it should be a common physical property of ferroelastics, not limited to ferroelectrics only, as demonstrated by the nonferroelectric but ferroelastic (GdI$_3$)$_2$Li monolayer here.

\subsection{Results of Mg-intercalated GdI$_3$}
Above results have demonstrated some interesting physical properties of half-doped GdI$_3$, except the expected high-temperature ferromagnetism. Then a natural question is that whether the one electron intercalation can make GdI$_3$ more like GdI$_2$, namely to be a FM semiconductor? To answer this question, the Mg intercalation is studied.

\begin{figure}
\includegraphics[width=0.48\textwidth]{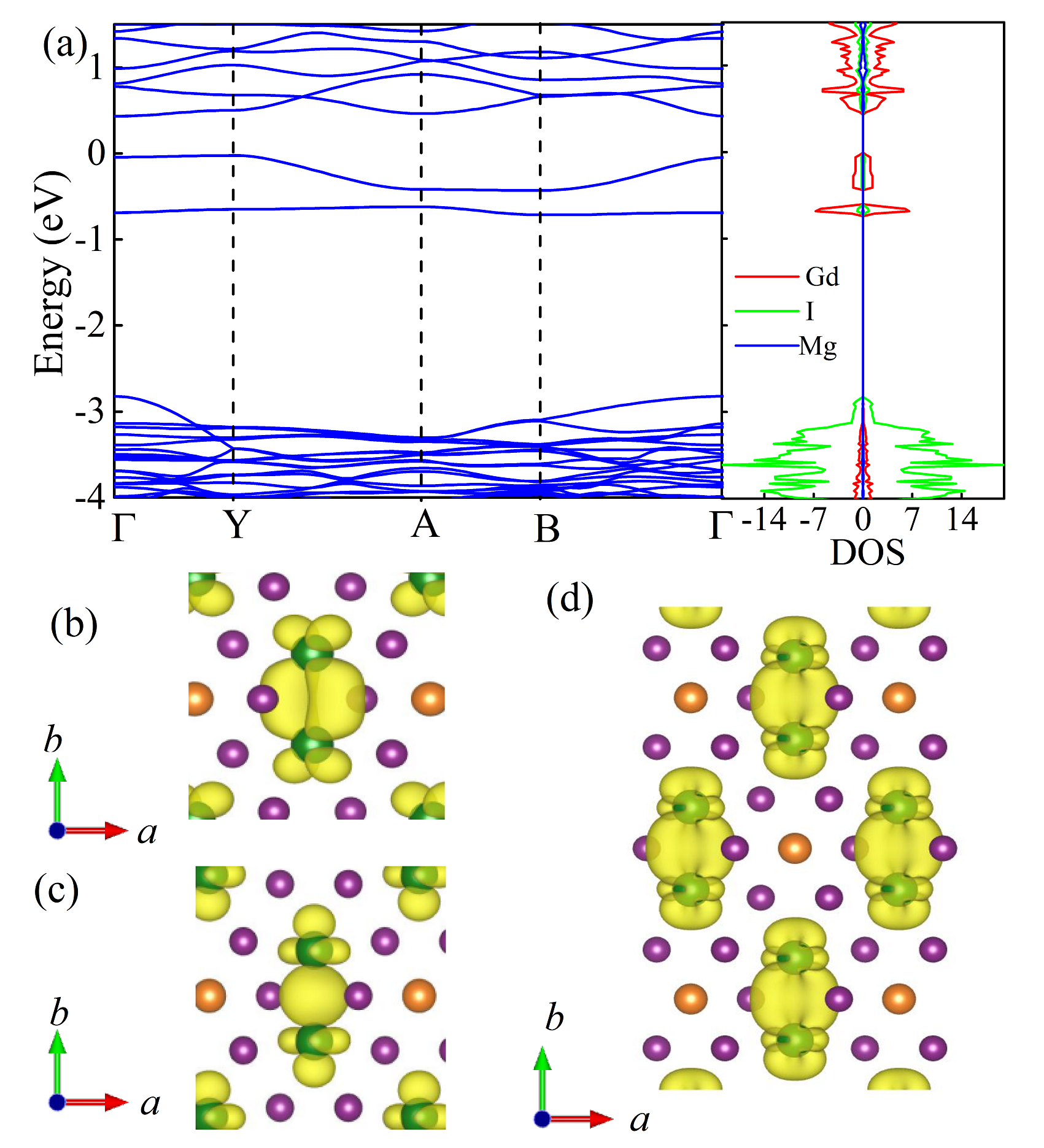}
\caption{Results of (GdI$_3$)$_2$Mg. (a) Electronic structure of Stripy-AFM state. The system remains a semiconductor, with two lowered $5d$ bands. Left: band structure; right: DOS. (b-d) Distribution of electrons. (b) and (c) are for two occupied $5d$ bands respectively, and (d) are the total occupation.}
\label{fig4}
\end{figure}

In (GdI$_3$)$_2$Mg, the valence of Gd becomes $+2$, identical to that in GdI$_2$. However, the Stripy AFM state remains the lowest energy one, according to our DFT calculations, as compared in Table~\ref{table3}. The magnetic moment becomes $\sim8$ $\mu_{\rm B}$/Gd, as expected.The MAE coefficient $K$ for (GdI$_3$)$_2$Mg is estimated as $1.05$ meV, which also prefers the in-plane alignment due to the SOC of Gd's $5d$ orbitals. The ferroelastic distortion is also prominent. In this sense, the Peierls transition remains the dominent factor, which opens a band gap at the Fermi level, as shown in Fig.~\ref{fig4}(a) (and Fig.~S4 in SM \cite{SM}).

How can the Peierls transition occur in the case of one electron doping? It is due to the multiple orbitals of Gd's $t_{2g}$ sector, which can separate one electron into half-occupation of two orbitals. Then the orbital-selective Peierls transition occurs (see Fig.~S5 in SM \cite{SM}), similar to the orbital-selective Mott transition in iron-based superconductors \cite{Yu:Prb}. The electron cloud shapes of two selected orbitals after Peierls transition are visualized in Figs.~\ref{fig4}(b-c) respectively, and their summation is shown in Fig.~\ref{fig4}(d). Correspondingly, the $b/a$ ratio is only $1.576$, $9\%$ shorter than the original $1.732$ of high symmetric one. In other words, the ferroelasticity is superior in Mg-intercalated case. In short, the physical effects of Mg intercalation is qualitatively similar to the Li-intercalation, due to the large electron capacity of multiple orbitals.

\begin{table}
\caption{Optimized structures of (GdI$_3$)$_2$Mg with different magnetic orders. Lattice constants ($a$ and $b$) and nearest-neighbor Gd-Gd distances ($d_l$: longer one; $d_s$: shorter one) are in unit of \AA. The energies are in unit of meV/f.u. and the FM one with high symmetric structure is taken as the reference.}
\begin{tabular*}{0.48\textwidth}{@{\extracolsep{\fill}}lcccccc}
\hline \hline
Order & Space group & Energy & $a$ & $b$ & $d_l$ & $d_s$ \\
\hline
FM & $P\bar{3}1m$ & 0 & 7.461 & - & 4.307 & - \\
N\'eel & $P\bar{3}1m$ & 139.9 & 7.511 & - & 4.336 & - \\
Zigzag & $C2/m$ & -43.8 & 7.168 & 15.160 & 4.938 & 3.985\\
Stripy & $C2/m$ & -129.3 &  7.877 & 12.416 & 4.834 & 3.404  \\
\hline \hline
\end{tabular*}
\label{table3}
\end{table}

Finally, it should be noted that the electron doping effects should not be seriously dependent on the position of intercalation, since here they are only electron donors. Other possible intercalation sites, if exist, will not alter the emergent results revealed above. Experimentally, the intercalation or adsorption is a common experimental technique in low-dimensional materials studies \cite{zhang:Nat.Commun,yan:PRX}.

In conclusion, a new 2D $f$-electron magnet GdI$_3$ and its intercalated variants have been theoretically studied, which exhibit interesting physical properties. First, the pristine GdI$_3$ monolayer can be easily exfoliated from its vdW bulk. Its magnetic ground state is the simplest N\'eel-type antiferromagnetism. Second, the half- or one-electron doping can be achieved via Li or Mg intercalation, which drives significant Peierls transition and changes the magnetic ground state. Other emergent physical issues, including ferroelasticity, multiferroicity, as well as magnetoresistivity, are predicted, all of which are expected to be experimentally verified in near future.

\begin{acknowledgments}
This work was supported by the National Natural Science Foundation of China (Grant No. 11834002). We thank the Tianhe-II of the National Supercomputer Center in Guangzhou (NSCC-GZ) and the Big Data Center of Southeast University for providing the facility support on the numerical calculations.
\end{acknowledgments}

\bibliography{ref}
\bibliographystyle{apsrev4-2}
\end{document}